\documentclass[9pt,twocolumn,twoside]{osajnl}
\journal{ol}
\setboolean{shortarticle}{true}
\usepackage{color}
\usepackage{amsmath}
\usepackage{ulem}
\usepackage{gensymb}

\newcommand{\be}{\begin{equation}}
\newcommand{\ee}{\end{equation}}
\newcommand{\beq}{\begin{eqnarray}}
\newcommand{\eeq}{\end{eqnarray}}
\newcommand{\bea}{\begin{array}}
\newcommand{\eea}{\end{array}}
 \usepackage{lipsum}
\usepackage{mathtools,amssymb,lipsum}
\usepackage{graphicx}
\usepackage{xcolor,ulem} 

\title{Enantioselection and chiral sorting of single microspheres using optical pulling forces}

\author[1,*]{Rfaqat Ali}
\author[2]{R. S. Dutra}
\author[3]{F. A. Pinheiro}
\author[3]{P. A. Maia Neto}
\affil[1]{Photonics Research Center, Applied Physics Department, Gleb Wataghin Physics Institute, P.O. Box 6165,
University of Campinas - UNICAMP, 13083-970 Campinas, SP, Brazil}
\affil[2] {LISComp-IFRJ, Instituto Federal de Educa\c{c}\~ao, Ci\^encia e Tecnologia, Rua Sebasti\~ao de Lacerda, Paracambi, RJ, 26600-000, Brasil}
\affil[3] {Instituto de F\'isica, Universidade Federal do Rio de Janeiro, Caixa Postal 68528, Rio de Janeiro, RJ, 21941-972, Brasil}

\affil[*]{Corresponding author: rali.physicist@gmail.com}

\begin{abstract}
We put forward a novel, twofold scheme that enables at the same time all-optical enantioselection and sorting of single 
multipolar chiral microspheres based on optical pulling forces exerted by two non-collinear, non-structured,
circularly-polarized light  sources. Our chiral resolution method can be externally controlled by varying the angle between their incident wavevectors,
allowing for a fine-tuning of the range of chiral indices for enantioselection. Enantioselectivity is achieved by choosing angles such that only particles with the same 
handedness of the light sources are pulled.
This proposal allows one to achieve all-optical sorting of  chiral microspheres with arbitrarily small chiral parameters, thus outperforming current optical methods.  
\end{abstract}

\setboolean{displaycopyright}{true}

\begin{document}

\maketitle

The concept of chirality pervades the natural world and occurs at all length scales, from the subatomic to the galactic. It is a geometrical property associated to non-superimposable mirror images, such as a collection of points, molecules, and nanostructures  ~\cite{Wagniere,Fan2010,Fan2012}.
The identification, sorting, and enantioselection of chiral substances consist of a very active, interdisciplinary, and relevant research topic of both scientific and technological interest \cite{zhang2005,cpreview2019,spivak2009,Ali2020d}

The advent of new optical methods and materials has fostered the development of new optical enantioselective techniques. Proposed methods include
 the use of structured light that deflects each enantiomer in opposite directions \cite{Kun2014,bradshaw2014,Canaguier2013,Cameron2014,Durand2015}, the enantioselective trapping via optical tweezers with single focused beams  \cite{Ali2020d,Ali2020a,Ali2020c,zhao2016,Patti2019}, and the application of lateral forces \cite{wang2014,hayat2015,Zhu2020}.
These lateral forces are dependent upon the particle chirality and are typically induced by linearly polarized beams, which separate particles with opposite handedness~\cite{wang2014,hayat2015,chen2016,zhan2017,Ho2017,Cao2017,Solomom2018}. However, most of the proposed enantioselective schemes based upon optical lateral forces are restricted to the dipolar or geometric optics regime~\cite{tkachenko2014,kravets2019,Tkachenko2014}. In addition, the existing optical methods for chiral resolution cannot pull and/or push particles of size comparable to the wavelength, where Mie theory applies and the enantioselective mechanism and methods are quite different.

To circumvent these limitations we put forward a novel enantioselective method for chiral particles with sizes comparable to the incident radiation, composed of two non-colinear, non-structured plane waves, using pulling optical forces. These forces pull a particle all the way towards the source without an equilibrium point as a result of a backward scattering force due to the interference of radiation multipoles, with many applications~\cite{chen2011,li2019,li2020,Brzobohaty2013,shi2020,Ding2019,Ali2020b,Shvedov2014}. By calculating the optical force acting on a chiral sphere beyond the dipolar approximation using Mie theory, we demonstrate that it strongly depends on the chiral parameter $\kappa$ and radius, allowing for enantioselection and chiral sorting. We also show that the optical force changes sign as a function of the angle between the two incident wave vectors, from pushing to pulling, demonstrating that one can externally control the enantioselective mechanism.  We show that this mechanism can selectively resolve particles with arbitrarily small chiral indexes 
by suitably adjusting the angle between the two propagation directions. 
This is in contrast to recent proposals that demonstrate enantioselection and chiral sorting 
of particles of sizes of the order of wavelength, in which the chiral parameter must be sufficiently
 large $\kappa\gtrsim 0.1$ thus restricting the range of potential applications~\cite{shi2020}.

\begin{figure} 
\includegraphics[width = 3.3in]{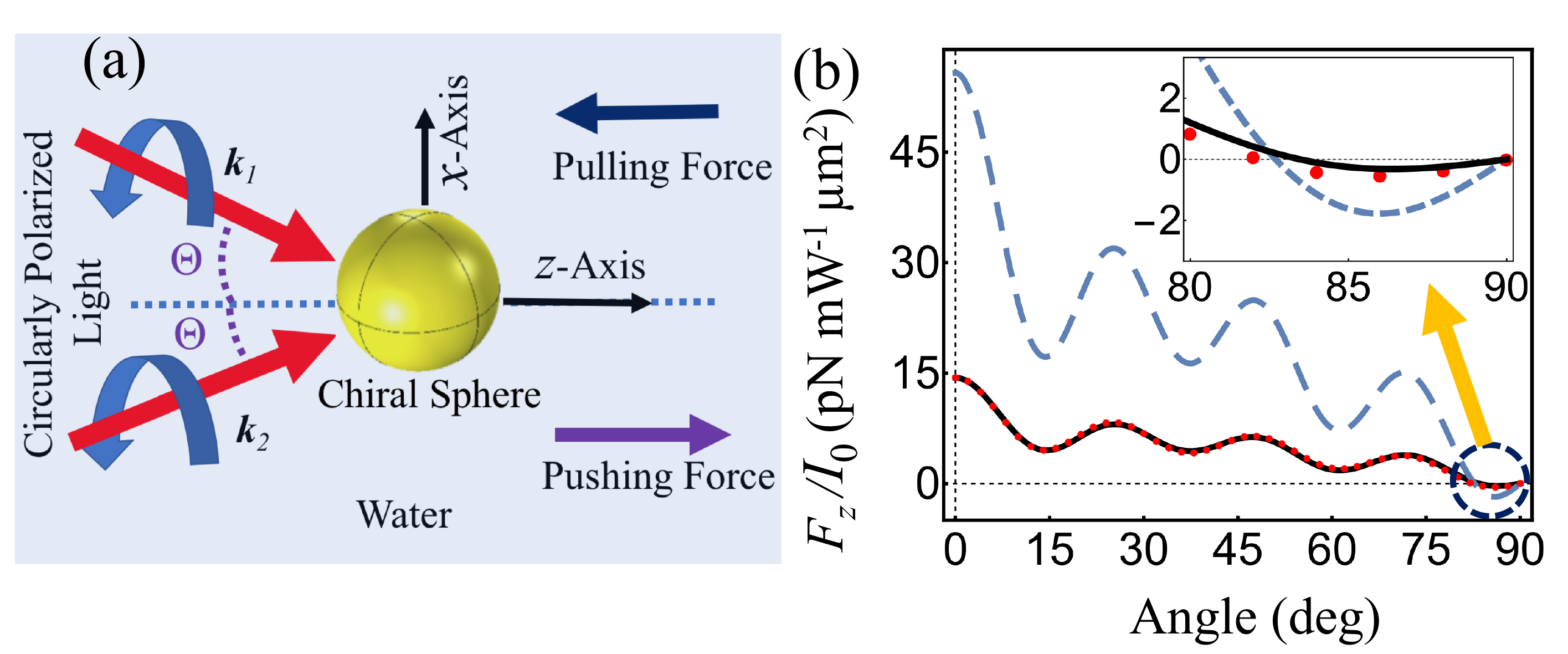}
\caption{ (a) Schematic diagram of a particle illuminated by two non-colinear circularly polarized { laser beams making an angle  $\Theta$  with respect to the $z$ axis. (b) Optical force acting on a chiral particle
with chiral parameter $\kappa=-0.3$ 
 as function of $\Theta$ in the case of right circular (dash), transverse electric (solid) and transverse magnetic (points) polarizations. }}  
\label{F1}
\end{figure}

We consider two circularly polarized laser beams (helicity $\sigma_j=\pm 1 $) positioned at an angle $\Theta$  with respect to the common-axis (z-axis) propagating in a non absorbing, non-magnetic  medium,  {as illustrated by Fig \ref{F1}(a).}
We model the (paraxial) laser beams as plane waves as their waists are typically much larger than the particle size, which we consider to be in the micrometer range.
The total electric field incident on the particle
 can then be expressed as the superposition
\begin{equation}
\mathbf{E}_{\rm in}=\sum^2_{j=1}E_0e^{i \sigma_{j} \phi_j} (\hat{\theta}_{j}+ i\sigma_{j}\hat{\phi}_{j})e^{i \mathbf{k}_{j}\cdot \mathbf{r}-\omega t}. \label{incidentfield}
\end{equation}
The wave vectors $\mathbf{k}_{j}({\theta}_{j}=\Theta, \phi_j)$ are such that propagation is on
the $xz$ plane. Hence we take  $\phi_{1}=0$ and $\phi_{2}=\pi.$ The beams share the same amplitude $E_0$ and
angular frequency  $\omega:$
 $\mathbf{k}_1=n_h\frac{\omega}{c}\{ \sin (\Theta ),\,0,\,  \cos (\Theta )\}$ and $\mathbf{k}_2=n_h \frac{\omega}{c}\{- \sin (\Theta ),\,0,\, \cos (\Theta )\}$, where $n_h$ is  the  refractive index of the host medium and $c$ is the speed of light.

{ We solve the Mie scattering by the spherical particle by expanding the electromagnetic fields in terms of Debye potentials $\Pi^E$ and $\Pi^M$ for electric and magnetic multipoles, respectively~\cite{Bowkamp1954}: 
\begin{eqnarray}
\Pi^{E} ({\bf r})= \sum^{\infty}_{\ell=1}\frac{({\bf r}\cdot {\bf E})_{}}{{\ell}({\ell}+1)}; \, \, \, \, \, \, \, \, \, \, \, \, \, \, \, \,
\Pi^{M}({\bf r}) =\sum^{\infty}_{\ell=1}\frac{({\bf r}\cdot {\bf H})}{{\ell}({\ell}+1)} \, .  \nonumber
\end{eqnarray}}

 { We  first expand the incident  field~(\ref{incidentfield}) into spherical multipolar waves and derive the {incident}  Debye potentials
 $\Pi^{E,{\rm in}}$ and $\Pi^{M,{\rm in}}$ as detailed in Ref.~\cite{Ali2020b}.  
The Wigner rotation matrix elements~ \cite{Edmonds1957}
 $d^{\ell}_{m,\sigma}(\Theta)$ 
  allow us to derive the potentials for an arbitrary direction of incidence  from 
 the simpler case of propagation along the $z$-axis. }
 The potentials for the scattered field $\Pi^{E,s}$ and magnetic  $\Pi^{M,s}$ are then obtained
  from the boundary conditions at the spherical surface.
 The explicit expressions { of $\Pi^{E,s}$ and $\Pi^{M,s}$ are written as partial-wave} sums over $\ell$ (for the total angular momentum $J^2$) and $m$ (corresponding to $J_z$) of the form
 $
\sum_{\ell,m}\equiv  \sum^{\infty}_{\ell=1}\, \sum^{\ell}_{m=-\ell}:
 $ 

\begin{equation}
\Pi^{E,s}= -\frac{E_{0}}{k}e^{- i \omega t}\sum_{\ell,m} i \gamma^{E}_{\ell m}\, h^{(1)}_{\ell}(kr)Y_{\ell,m}(\theta,\phi),
\end{equation}

\begin{equation}
\Pi^{M,s}= -\frac{H_{0}}{k}e^{- i \omega t}\sum_{\ell,m} \gamma^{M}_{\ell m}\, h^{(1)}_{\ell}(kr)Y_{\ell,m}(\theta,\phi),
\end{equation}
where  $H_0=\sqrt{\epsilon_h\epsilon_0/\mu_0}\,E_0,$ $k=n_h\frac{\omega}{c}$, { $\epsilon_0$   and $\mu_0$ are the vacuum permittivity and permeability, respectively, }
and $ h^{(1)}_{\ell}$ and $Y_{\ell,m }$ denote the
 spherical Hankel functions and the spherical harmonics, respectively~\cite{DLMF25.12}. In addition,

\[
\gamma^{E}_{\ell m}=\sqrt{\frac{4\pi(2\ell+1)}{\ell(\ell+1)}}(i)^\ell [\sigma_{1}A^{(\sigma_1)}_{\ell} d^{\ell}_{m,\sigma_{1}}(\Theta)-(-1)^m\sigma_{2}A^{(\sigma_2)}_{\ell} d^{\ell}_{m,\sigma_{2}}(\Theta)]
\]

\[
\gamma^{M}_{\ell m}=\sqrt{\frac{4\pi(2\ell+1)}{\ell(\ell+1)}}(i)^\ell  [B^{(\sigma_1)}_{\ell} d^{\ell}_{m,\sigma_{1}}(\Theta)-(-1)^mB^{(\sigma_2)}_{\ell} d^{\ell}_{m,\sigma_{2}}(\Theta)].
\]
 From now on, we assume that both beams have the same helicity:
   $\sigma_1=\sigma_2=\sigma.$

The coefficients   $A^{(\sigma)}_{\ell}$ and $B^{(\sigma)}_{\ell}$  are  the Mie coefficients of the equivalent achiral sphere scattering a circularly polarized beam of helicity $\sigma$, that appear as~\cite{Ali2020c}
\begin{eqnarray}
 A^{(\sigma)}_{\ell} \leftrightarrow a_{\ell}+ \,  i \,\sigma d_{\ell}; \label{equivalence}
  \; \;  \; \;  \; \;  \; \;  \; \;  B^{(\sigma)}_{\ell} \leftrightarrow b_{\ell} -  i\sigma  \, c_{\ell}
\label{equivalence2}
\end{eqnarray} 
where $a_{\ell}$,\, $b_{\ell}$,\, $c_{\ell}$ and $d_{\ell}$ are the Mie coefficients for the homogeneous chiral sphere as defined in Ref.~\cite{Ali2020c}. When we take $\kappa$ to zero ($c_l=d_l=0$) we recover the Debye potentials for an achiral sphere \cite{Ali2020b}. 
Equation (\ref{equivalence}) shows that one can map the problem of light scattering by a homogeneous chiral sphere into the simpler original Mie scattering problem for achiral particles \cite{Bohren1974}.

{ The total electromagnetic fields are then expressed in terms  of the potentials
$\Pi^{E} = \Pi^{E,in} +  \Pi^{E,s}$ and  $\Pi^{M} = \Pi^{M,in} + \Pi^{M,s}$  as 
\begin{eqnarray}
\label{EfromPi}
{\bf E}&=  & \nabla \times \nabla \times \left({\bf r}\, \Pi^{E}\right) + i\omega \mu_0 \,  \nabla \times\left({\bf r}\, \Pi^{M}\right)\\
\label{HfromPi}
{\bf H}&=&  \nabla \times \nabla \times\left({\bf r}\, \Pi^{M}\right) - i\omega \epsilon_h\epsilon_0\,  \nabla \times\left({\bf r}\, \Pi^{E}\right) \, ,
\end{eqnarray} }
 Finally, we  determine the force using the Maxwell stress tensor:
\begin{equation}
{\bf F}= \lim_{r\rightarrow \infty}\left[ -\frac{r}{2} \int {\bf r}\left( \epsilon_h \epsilon_0 E^2 + \mu_0 H^2 \right) d\Omega  \right],
\label{Max}
\end{equation}
Since the incident plane waves have the same amplitude and polarization, the  optical force will point along the $z$ axis by symmetry. 

As the stress tensor is quadratic in the  total electric  $\textbf{E}$ and magnetic $\textbf{H}$ fields, the optical force  
{ $F_z= F_{\rm s,z}+F_{\rm e,z}$}  has two distinct contributions:
the extinction term  $F_{\rm e,z}$ arises from cross terms of the form ${\bf E}_{\rm in}\cdot \textbf{E}^*_{\rm s}$ (and likewise for the magnetic field) and represents  the rate of linear momentum removal from the incident fields.
Part of this momentum is carried away by the scattered fields, 
and the rest is transferred to the particle as an optical force. Thus, the second contribution to the optical force $F_{\rm s,z},$
which is  quadratic in  $\textbf{E}_s$ and   $\textbf{H}_s,$ represents the negative of the rate of momentum contained in the scattered electromagnetic fields.
 We find 
\[
 F_{z,\rm s}= \frac{\epsilon_h \epsilon_0 E_{0}^2}{k^2}\sum_{\ell,m}{\rm Im}\,\biggl\lbrace \ell(\ell+2) \sqrt{\frac{(\ell+1-m)(\ell+1+m)}{(2\ell+1)(2\ell+3)}}\]
\begin{equation}
\times \biggl[ 
 \gamma^{{\rm{E}}}_{\scriptscriptstyle{\ell,m}}\gamma^{{\rm{E}}^*}_{\scriptscriptstyle{\ell+1,m}}
+ 
\gamma^{{\rm{M}}}_{\scriptscriptstyle{\ell,m}}\gamma^{{\rm{M}}^*}_{\scriptscriptstyle{\ell+1,m}}\biggr]
-im \gamma^{{\rm{M}}}_{\scriptscriptstyle{\ell,m}}\gamma^{{\rm{E}}^*}_{\scriptscriptstyle{\ell,m}}\biggr\rbrace . \label{Fzs}
\end{equation}
\[
F_{z,e}=\frac{2\pi\epsilon_h \epsilon_0 E_{0}^2}{k^2}\,\sum_{\ell,m}
\sqrt{\frac{\ell(\ell+1)(2\ell+1)}{4\pi}}{\rm Re} \biggl[\biggl(\sigma_{1}\gamma^{{\rm{E}}^*}_{\scriptscriptstyle{\ell,m}}
+\gamma^{{\rm{M}}^*}_{\scriptscriptstyle{\ell,m}} \biggr)G^{C1}_{\ell,m}
\]
\begin{equation}
- \, 
\biggl(\sigma_{2}\gamma^{{\rm{E}}^*}_{\scriptscriptstyle{\ell,m}}
+\gamma^{{\rm{M}}^*}_{\scriptscriptstyle{\ell,m}} \biggr)G^{C2}_{\ell,m} \biggr], \label{Fze}
\end{equation}
We have defined the coefficients
\begin{equation}
G_{\ell,m}^{C1}= (i)^\ell \cos\Theta\,  d^{\ell}_{m,\sigma_{1}}(\Theta); \, \, \, \, \, \, \, \, \, 
G_{\ell,m}^{C2}= (i)^\ell(-1)^m \cos\Theta \, d^{\ell}_{m,\sigma_{2}}(\Theta).
\end{equation}

 We consider that the laser beams have wavelength $\lambda_{0}=1064\, {\rm nm}.$  They impinge on a chiral microsphere with  relative permittivity  $\epsilon_s=2.5$ and chiral parameter $\kappa$ ranging from $-0.5$ to $+0.5.$  The host medium is an aqueous  solution with refractive index $n_h= 1.332$. In the following discussion, we normalize the optical force by the intensity $I_0 {=\sqrt{\epsilon_h\epsilon_0/\mu_0}E_0^2/2}$ of each incident beam.

\begin{figure}
 \centering
\includegraphics[width = 3.4 in]{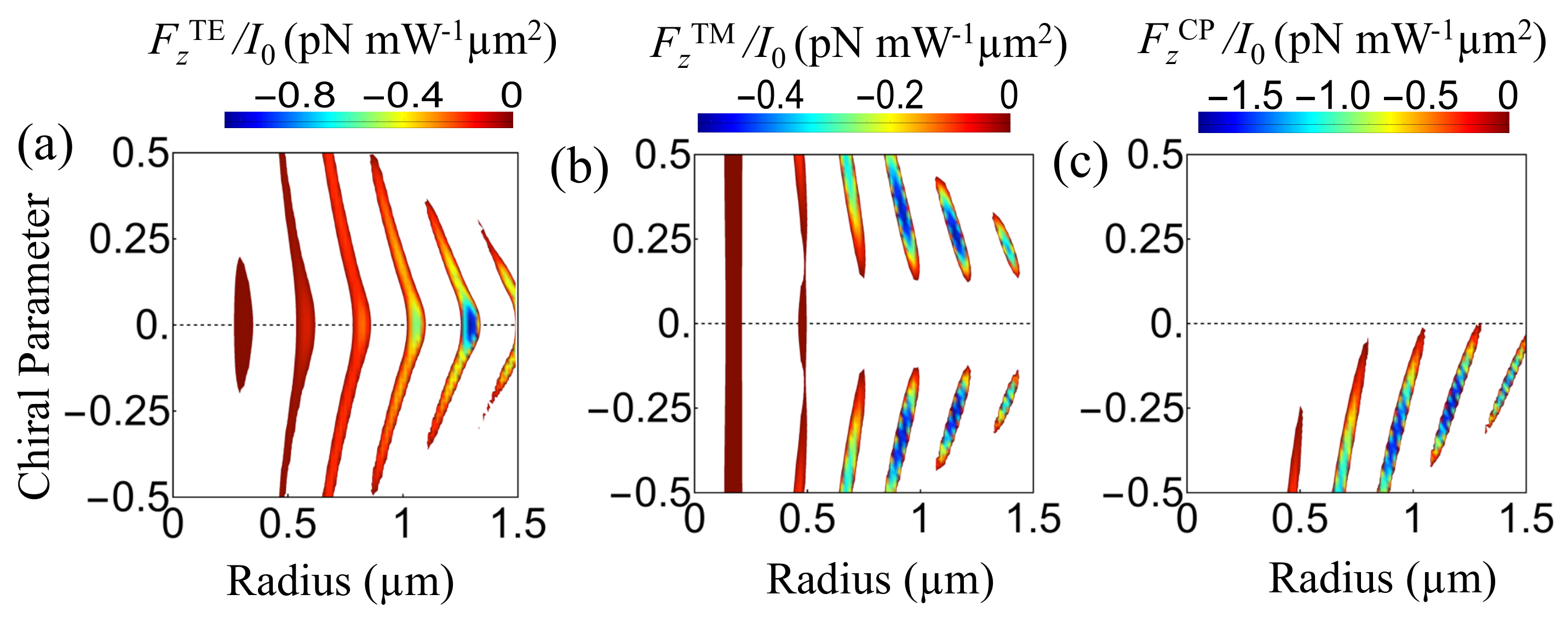}\\
\caption{Normalized optical force acting on a chiral particle
as a function of radius and chiral parameter.  Regions where 
$F_z<0$ (pulling) are indicated
by color patches in the density plot.
The particle is under illumination by two (a) transverse electric,  (b)  transverse magnetic,  and (c)  right circularly polarized laser beams of wavelength $\lambda_0=1064\, {\rm nm}$. The half-angle between the wave vectors is $\Theta = 85\degree.$  } 
\label{F2}
\end{figure}

In addition to the circular polarization taken in Eq.~(\ref{incidentfield}), we also consider linearly-polarized beams 
where either the electric (TE) or magnetic (TM) fields are transverse to the plane defined by the wave vector.
In Fig. \ref{F1}(b), we plot the axial force $F_z$ acting on a  microsphere of radius $930\,{\rm nm}$ and chiral parameter $\kappa=-0.3$ as a function of $\Theta$  in the case of 
TE (solid) and TM (points) linear polarizations, as well as for 
right circular polarization (RCP, dash), which corresponds to $\sigma=-1.$
Figure \ref{F1}(b) reveals that the axial force is positive (pushing) for $ \Theta \lesssim 82\degree$, exhibiting an oscillatory behaviour as a function of $\Theta$ and eventually changes sign for $82\degree \lesssim \Theta < 90\degree$ for all polarisation states. At a large incident angle, the  interference between them leads to 
maximally  scattered  momentum along the  forward direction,  which is the reason for a strong recoiling force along backward direction.
This crossover from pushing (force along the propagation direction) to pulling  forces can be externally controlled by varying $\Theta$ and results from the excitation and interference between electric and magnetic multipoles. 
 It is worth mentioning that circularly-polarized plane waves exert a stronger overall force in comparison to the case of linear polarization.

 In Fig. \ref{F2}, the optical force is calculated as a function of both the chirality parameter $\kappa$ and sphere radius for $\Theta = 85\degree.$
 We consider (a) TE, (b) TM and (c) RCP waves. The regions in color indicate the parameters yielding an optical pulling force while white regions correspond to pushing forces.
It is clearly demonstrated in Fig. \ref{F2}(a) and Fig. \ref{F2}(b) that the linearly polarized waves exert identical forces on both enantiomers with opposite chirality parameters $\kappa$,  as expected from symmetry. { In the case of TE polarization, pulling takes place regardless of chirality in several size intervals~\cite{Ali2020b}, but the specific pulling intervals depend slightly on $|\kappa|.$}

On the other hand, RCP light interacts differently with each enantiomer and exerts a chirality-dependent optical force as illustrated by Fig. \ref{F2}(c).
This is corroborated by the analytical result in dipolar limit, which is derived by taking $\ell =1$ in equations 
 \ref{Fzs} and \ref{Fze} (see Ref. \cite{Ali2020b} for a similar derivation for achiral microspheres).
Therefore, the pulling force can be applied to enantioselection of chiral particles provided that circularly polarized plane waves are used.
One can choose the chirality of the particles that are optically pulled
 by selecting the helicity of the plane waves with the same handedness.  Fig.~\ref{F2}(c) shows that right-handed particles ($\kappa<0$) are 
 pulled by using RCP laser beam. 
 In addition, one can also pull left-handed particles ($\kappa>0$) by selecting LCP laser beams instead. 
 Indeed, since the optical force is invariant when changing both $\kappa$ and $\sigma,$
 the density plot in the case of left circular polarization (LCP, $\sigma=+1$) is the mirror image of Fig. \ref{F2}(c) after taking $\kappa\rightarrow -\kappa.$

 Figure \ref{F2}(c) shows that the condition for optical pulling  is also governed by the microsphere size. Hence the pulling force sorts particles according to their 
 radii within a range that becomes increasingly correlated with the value of $\kappa$ as the size increases.

\begin{figure} 
\includegraphics[width = 3.5in]{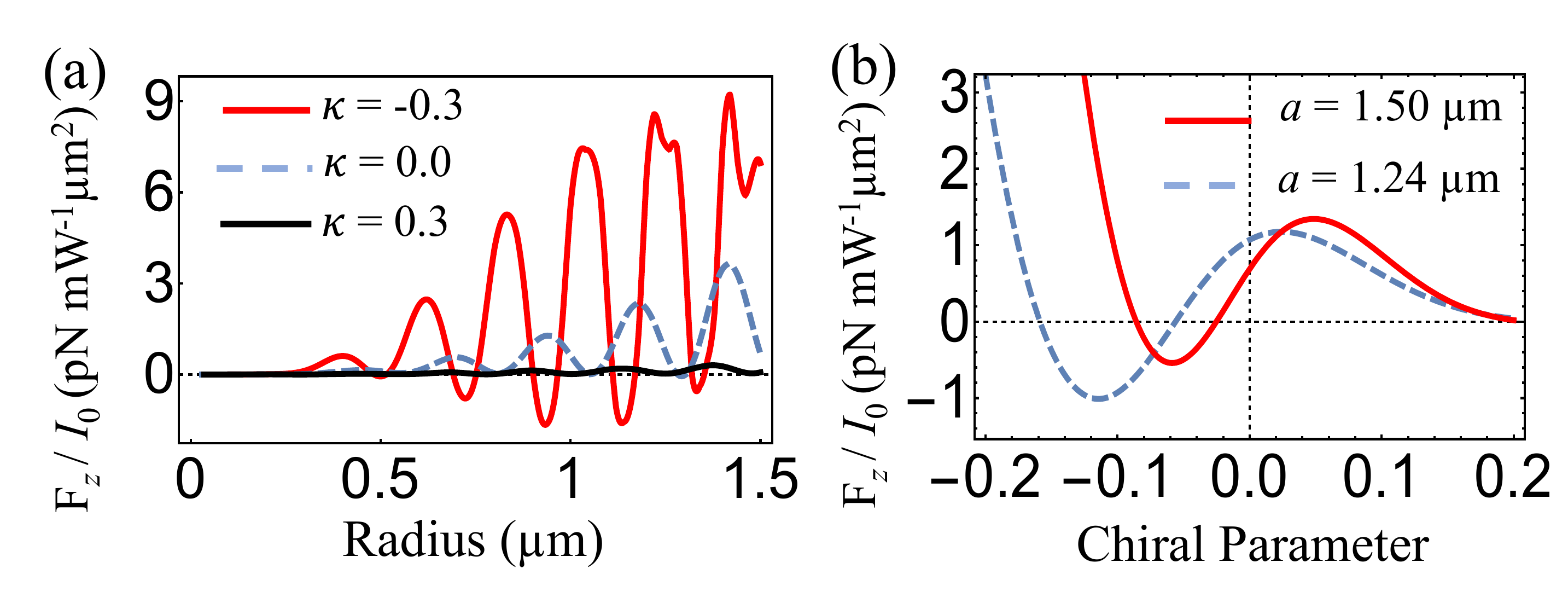}
\caption{  Optical force acting on chiral particle by two RCP laser beams, (a) as a function of the microsphere radius for chiral parameters
 $\kappa=-0.3$ (red), $\kappa=0.$ (dash) and  $\kappa=0.3$ (black);
(b) as a function of  $\kappa$
for radii $1.24\,\mu {\rm m}$ (dash) and $1.50\,\mu {\rm m}$ (solid). The half-angle between the laser beams is fixed at $\Theta=85 \degree$ for all curves.}
\label{F3}
\end{figure}

The panels of Fig.~\ref{F2} indicate an overall increase in the 
 magnitude of the pulling force as one departs from the dipolar limit by increasing
the particle size, because a larger number of electric and magnetic multipoles come into play.
When considered as a function of radius, the optical force oscillates with an increasing amplitude, as shown in
 Fig.~\ref{F3}(a) for 
 $\kappa=-0.3$ (red), $\kappa=0$ (dash)  and $\kappa=0.3$ (black).
 In both panels of Fig.~\ref{F3}, we take 
 RCP beams making a half-angle  $ \Theta=85\degree.$  
 The magnitude of the pulling force peaks
 in the intermediate size range defined by the value of the laser wavelength
 $\lambda_0=1.064\,\mu{\rm m}.$ As the radius increases further, too many multipoles contribute to the force, making it harder to  identify the condition of constructive interference along the 
 forward direction.

 Figure~\ref{F3}(b) provides additional information on how the magnitude of the pulling effect decreases as the radius increases past $\lambda_0.$ 
We plot the optical force variation with $\kappa$ and show
that the maximum pulling force for a radius of $1.50\,\mu{\rm m}$ (solid) is indeed smaller than for a radius of $1.24\,\mu{\rm m}$ (dash).
More importantly, Fig.~\ref{F3}(b) illustrates the correlation between size and chirality already discussed in connection with  Fig.~\ref{F2}(c).
Indeed, the interval of chiral parameters selected by the pulling effect clearly depends on the radius. In the example shown in the figure, 
particles with radius $1.50\,\mu{\rm m}$ are pulled for smaller values of $|\kappa|,$ with the handedness selected by the helicity of the laser beams.

\begin{figure} 
\includegraphics[width = 3.3in]{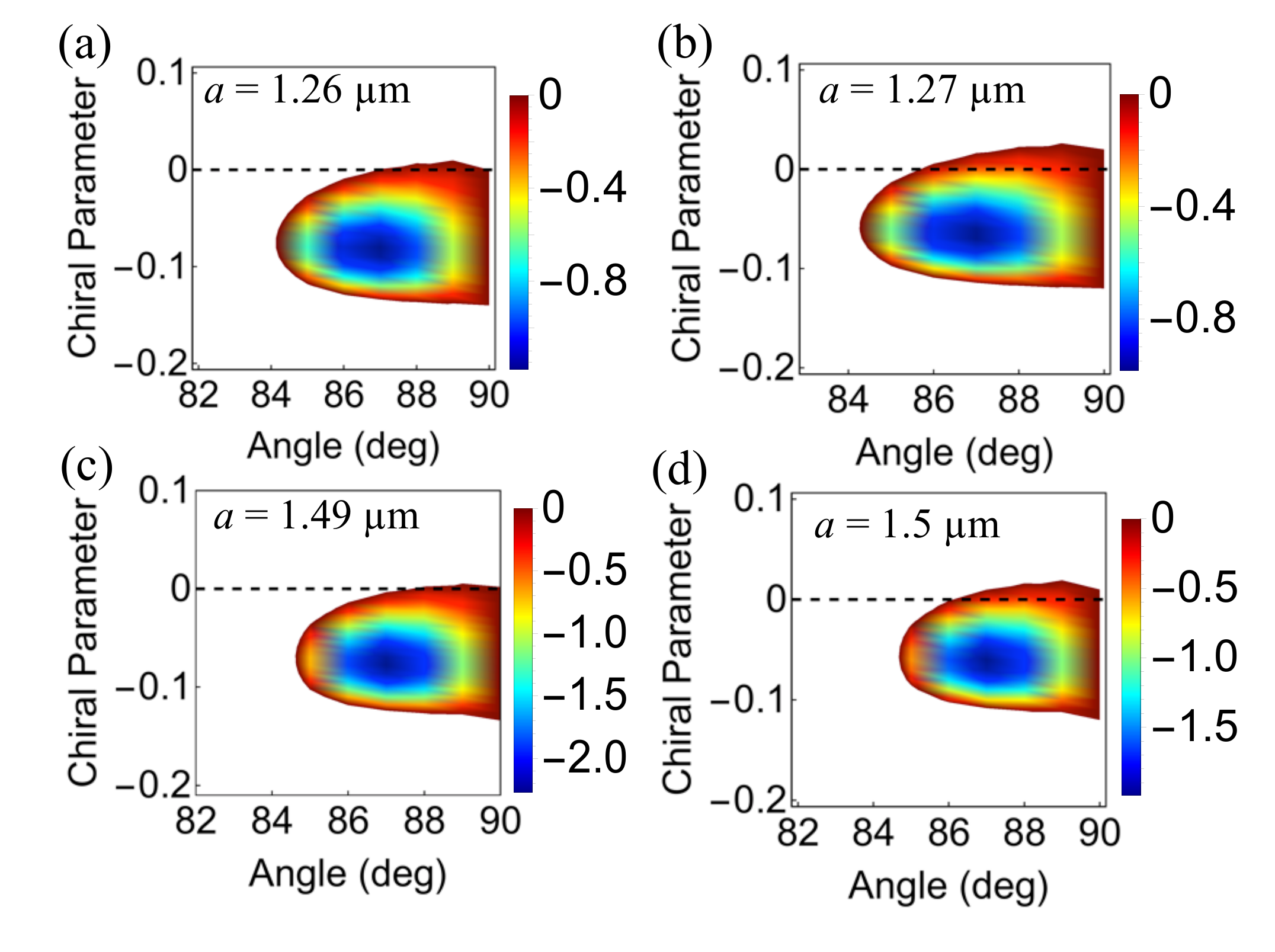}
\caption{ Normalized optical force acting on a chiral particle
as a function of chiral parameter and half-angle between the incident laser beams, which are both right circularly polarized. 
Regions where $F_z<0$ (pulling) are indicated
by color patches in the density plot.
The microsphere radius is (a) $1.26\,\mu{\rm m}$, (b) $1.27\,\mu{\rm m}$, (c) $1.49\,\mu{\rm m}$ and  (d) $1.50\,\mu{\rm m}$. }  
\label{F4}
\end{figure}

 A sought-after scenario for applications is to sort particles with chirality parameters $\kappa$ within a selected range of values by controlling some external parameter. 
Since the range of values of $\kappa$ leading to pulling depends critically on the half-angle $\Theta,$ 
it is possible to sort particles with a desired chirality by adjusting the directions of the laser beams. 
In Fig.~\ref{F4}, we plot the normalized optical force $F_z/I_0$ as a function of $\Theta $ and $\kappa$ for RCP beams and microsphere radii 
(a) $1.26\,\mu{\rm m}$, (b) $1.27\,\mu{\rm m},$ (c) $1.49\,\mu{\rm m}$ and (d) $1.50\,\mu{\rm m}.$
As in Fig.~\ref{F2}, only pulling forces ($F_z<0$) are shown for clarity. 
In all cases, the minimum $|\kappa|$ leading to pulling depends strongly on $\Theta,$ and can be tuned to arbitrarily small values by 
selecting the appropriate angle for a given radius. For instance, for the radius $1.26\,\mu{\rm m},$ corresponding to panel (a), 
one can pull particles with chiral parameters as low as $\kappa\approx - 4\times 10^{-3}$ by selecting $\Theta=86.5\degree.$
Similar examples can be inferred for the other panels shown in the figure. 
 Such capability of sorting particles with unprecedented,  very small  chiral parameters clearly outperforms  existing chiral resolution methods based on optical forces~\cite{shi2020},
which are typically limited to chiral particles satisfying $|\kappa|\gtrsim 0.1.$  { Figure~\ref{F4} also reveals that the maximal angle $\Theta=90\degree$ does not impose a fundamental limit on the values of $\kappa$ of the particles that can be sorted. Indeed, these values depend on the radius, which hence can be exploited  as a free parameter to select the desired range of $\kappa$ for chiral resolution.}

Figure~\ref{F4} also indicates the maximum value of $\Theta$ 
that leads to pulling only of particles with the same handedness of the circularly-polarized laser beams. 
In the example shown in the figure, we have taken RCP beams, so pulling is achieved only for
right-handed particles
 ($\kappa<0$) as long as 
$\Theta$ is smaller than the value at the intersection between the colored region and the horizontal dashed lines in Fig.~\ref{F4}.
 
 In summary, we have put forward an  all-optical method for chiral resolution and enantioselection of single chiral microspheres 
based on optical pulling forces or tractor beams.  
Chiral particles with the same handedness of the 
circularly-polarized laser beams are optically pulled, whereas those with opposite handedness are flushed by radiation pressure. 
It is possible to fine-tune the
range of chiral parameters by adjusting the angle 
$2\Theta$ between the laser beams, which can be controlled with a high precision in typical optical setups. 
Our proposal allows for an all-optical sorting of multipolar chiral microspheres with arbitrarily small chiral parameters, thus outperforming current optical methods.  Altogether our findings pave the way for novel applications of optical pulling forces, such as chiral resolution, sorting, and self-assembling of chiral microparticles.

\subsection*{ Funding} 
Conselho Nacional de Desenvolvimento Cient\'{\i}fico e Tecnol\'ogico (CNPq),
Coordena\c c\~ao de Aperfei\c coamento de Pessoal de N\'{\i}vel Superior
 (CAPES),  Instituto Nacional de Ci\^encia e Tecnologia de Fluidos Complexos (INCT-FCx) ,
Fundac\~ao de Amparo a Pesquisa do Estado do Rio de Janeiro (FAPERJ) and
Fundac\~ao de Amparo a Pesquisa do Estado de S\~ao Paulo  (FAPESP) (2014/50983-3 and 2020/03131-2).

\subsection*{Acknowledgments} We thank S. Iqbal, G. Wiederhecker and F. S. S. da Rosa for inspiring discussions. 

\subsection*{ Disclosures} The authors declare no conflicts of interest.

\newpage

\end{document}